\documentclass[showpacs,aps,prb,reprint,twocolumn]{revtex4-1}

\usepackage{mathtools}
\usepackage{commath}
\usepackage{amsmath,bm}
\usepackage{graphicx}
\usepackage{hyperref}
\usepackage{txfonts}

\draft % marks overfull lines with a black rule on the right

\begin{document}

%\preprint{}

\title{Full Slonczewski-Weiss-McClure parametrization of few-layer twistronic graphene}
%\author{Author list}
%\author{Aitor Garcia-Ruiz${}^{1,2}$, Hai-Yao Deng${}^{2}$, Vladimir Enaldiev${}^{1,2}$, Vladimir I. Fal'ko${}^{1,2,4]$}
\author{Aitor Garcia-Ruiz$^{1,2}$, Haiyao Deng${}^{3}$, Vladimir V. Enaldiev${}^{1,2,4}$, Vladimir I. Fal'ko${}^{1,2,5}$}
\affiliation{$^{1}$National Graphene Institute, University of Manchester, Booth Street East, Manchester M13 9PL, UK}
\affiliation{$^{2}$Department of Physics and Astronomy, University of Manchester, Oxford Road,Manchester, M13 9PL, UK}
\affiliation{$^{3}$School of Physic and Astronomy, Cardiff University, 5 The Parade, Cardiff CF 24 3AA, Wales, UK}
\affiliation{$^{4}$Kotel'nikov Institute of Radio-engineering and Electronics, Russian Academy of Sciences, 11-7 Mokhovaya St, Moscow, 125009 Russia}
\affiliation{$^{5}$Henry Royce Institute for Advanced Materials, University of Manchester, Oxford Road, Manchester, M13 9PL, UK}
\begin{abstract}
We use a hybrid $k \cdot p$ theory - tight binding (HkpTB) model to describe interlayer coupling simultaneously in both Bernal and twisted graphene structures. For Bernal-aligned interfaces, HkpTB is parametrized using the full Slonczewski-Weiss-McClure (SWMcC) Hamiltonian of graphite \cite{McCann_2013}, which is then used to refine the commonly used minimal model for twisted interfaces\cite{Lopes_2007, Bistritzer_2011}, by deriving additional terms that reflect all details of the full SWMcC model of graphite. We find that these terms introduce some electron-hole asymmetry in the band structure of twisted bilayers, but in twistronic multilayer graphene, they produce only a subtle change of moir\'{e} miniband spectra, confirming the broad applicability of the minimal model for implementing the twisted interface coupling in such systems.

\end{abstract}

\maketitle

\section{introduction}
\label{sec:int}
The discovery of superconductivity in twisted bilayer graphene (tBLG)~\cite{Cao_2018-1,Cao_2018-2, Yankowitz_2019,Lu_2019,Cao_2020-1} has renewed interest in multilayer graphene with one \cite{Sharpe_2019, Polshyn_2020, Shi_2020, Chen_2019-1, Burg_2019, Shen_2020, Chen_2020, Cao_2020-2, Cao_2020-2, Liu_2020, Rickhaus_2020} or multiple \cite{Tsai_2020,Park_2021,Hao_2021} twisted interfaces, exploring topological and correlated-driven phases \cite{Sharpe_2019, Polshyn_2020, Shi_2020, Chen_2019-1, Burg_2019, Shen_2020, Chen_2020, Cao_2020-2, Cao_2020-2, Liu_2020, Rickhaus_2020}. Various theoretical approaches, including continuum model \cite{Lopes_2007, Bistritzer_2011, Koshino_2015, Lopes_2012, Mele_2011, Nguyen_2017, Moon_2012, Moon_2013}, tight-binding \cite{Moon_2012, Moon_2013, Morell_2010, Sboychakov_2015, Po_2019, Lin_2020, Carr_2019} and density functional theory (DFT)\cite{Carr_2019, Trambly_2010, Trambly_2012, Uchida_2014, Lucignano_2019}, have established a link between these phenomena and the existence of nearly flat bands for a twist angle $\theta\approx1.1^\circ$, known as ``magic angle"\cite{Tarnopolsky_2019}. These nearly dispersionless bands were theoretically predicted for twisted bilayers \cite{Bistritzer_2011} and recently observed with ARPES \cite{Utama_2021}. A detailed analysis in twistronic graphene structures involving few-layer graphene flakes requires refinement of theoretical models for their dispersion. In particular, one may wonder how well the use of the full set of interlayer couplings for bilayers or trilayers with Bernal stacking (which affect the miniband dispersion at low energies) comply with the minimal interlayer coupling Hamiltonian describing the hybridization across the twisted interface\cite{Lopes_2007,Bistritzer_2011}. 

To answer the above question, we develop a hybrid $k\cdot p$ tight-binding (HkpTB) model and relate its parameters to the Slonczewski-Weiss-McClure (SWMcC) Hamiltonian for graphite \cite{Slonczewski_1958, McClure_1957, McClure_1960}. The conventional minimal model \cite{Lopes_2007, Bistritzer_2011} is based on the two-centre approximation \cite{Lopes_2012, Koshino_2015} (TCA), which constructs the interlayer coupling integral from site-to-site hopping. In HkpTB approach, we express Bloch states as plane-waves localised in the vertical direction, with the interlayer coupling (in z-direction) implemented using the framework of tight-binding model. We derive additional terms in the twistronic graphene Hamiltonian which account for linear-in-momentum corrections, proportional to $v_3$ and $v_4$ SWMcC parameters, stacking-dependent on-site potentials, and next-nearest layer hoppings, $\gamma_2$ and $\gamma_5$.

The analysis below is structured as follows. In Sec.~\ref{sec:form}, we describe the approach, set notations and then apply it to bilayer graphene with an arbitrary lateral offset between the lattices of the two layers, which enables us to fully parametrise our calculations. In Sec.~\ref{sec:tblg}, we use that parametrization to derive the Hamiltonian for tBLG, highlighting corrections beyond the minimal model. In Sec.~\ref{sec:dispersion_tblg}, we compute the band structure of tBLG and discuss the influence of the correcting terms, and in Sec.~\ref{sec:dispersion_ttlg}, we analyse minibands in twisted trilayer (1+2) graphene.

\section{Formalism}
\label{sec:form}

First, we revisit the intralayer and interlayer coupling in a bilayer system using the basis of eigenfunctions, $\Psi_{l,\boldsymbol{k}}$, in each layer,
\begin{align}\label{eq:schrodinger_monolayer}
\left[
\frac{\hat{\boldsymbol{p}}^2}{2m_e}+
V_{l}(\boldsymbol{r},z)\right]|\Psi_{l,\boldsymbol{k}}\rangle=
E_w|\Psi_{l,\boldsymbol{k}}\rangle.
\end{align}
Here, $l=t,b$ labels the top and bottom layers, respectively, $E_w$ is the work function of the state $\Psi_{l,\boldsymbol{k}}$ in graphene, and $V_{l}(\boldsymbol{r},z)$ is the potential created by carbon atoms in the layer $l$. For a bilayer, this transforms into single-particle Schr\"{o}dinger equation,
\begin{align}
\left[
\frac{\hat{\boldsymbol{p}}^2}{2m_e}+
V_{t}(\boldsymbol{r},z)+
V_{b}(\boldsymbol{r},z)\right]
|\Psi\rangle=
E|\Psi\rangle,
\end{align}
which we solve using the basis of single-layer states \cite{Dresselhaus_2001}, $|\Psi_{\boldsymbol{k}}\rangle= \sum_{l} c_{l,\boldsymbol{k}} |\Psi_{l,\boldsymbol{k}}\rangle$, and find the bilayer band structure from the following matrix equation,
\begin{align}\label{eq:Matrix_Equation_general}
\hat{H}_{0}C=E(\mathbb{I}+\mathbb{S})C,
\end{align}
where $C$ is a column matrix made of coefficients $c_{l,\boldsymbol{k}}$ with different quasi-momentum and layer index, $\mathbb{I}$ is the unit matrix, and $\mathbb{S}$ contains the overlap integrals between pairs of eigenstates in the upper and lower layers,
\begin{equation}\label{eq:Overlap_general}
S(l\boldsymbol{k},l'\boldsymbol{k}')=
\langle \Psi_{l,\boldsymbol{k}}|
\Psi_{l',\boldsymbol{k}'}\rangle - \delta_{\boldsymbol{k},\boldsymbol{k}'} \delta_{l,l'}.
\end{equation}
The matrix elements of the operator $\hat{H}_0$ are
\begin{align}\label{eq:H0_intra_general}
H_0(l\boldsymbol{k},l\boldsymbol{k}')=&
\langle \Psi_{l,\boldsymbol{k}}
|\frac{\hat{\boldsymbol{p}}^2}{2m_e}+
V_{t}(\boldsymbol{r},z)+
V_{b}(\boldsymbol{r},z)|
\Psi_{l,\boldsymbol{k}'}\rangle=\\
=&\,E_w+
\langle \Psi_{l,\boldsymbol{k}}
|
V_{\bar{l}}(\boldsymbol{r},z)|
\Psi_{l,\boldsymbol{k}'}\rangle\nonumber,
\end{align}
where $\bar{l}$ refers to the opposite layer of $l$, and
\begin{subequations}\label{eq:H0_general}
\begin{align}
H_0(t\boldsymbol{k},b\boldsymbol{k}')\!=&
\langle \Psi_{t\boldsymbol{k}}
|\frac{\hat{\boldsymbol{p}}^2}{2m_e}\!+\!
V_{t}(\boldsymbol{r},z)\!+\!
\frac{\hat{\boldsymbol{p}}^2}{2m_e}\!+\!
V_{b}(\boldsymbol{r},z)\!-\!
\frac{\hat{\boldsymbol{p}}^2}{2m_e}|
\Psi_{b\boldsymbol{k}'}\rangle\nonumber\\
=&\,2E_wS(t\boldsymbol{k},b\boldsymbol{k}')-
\langle \Psi_{t\boldsymbol{k}}
|\frac{\hat{\boldsymbol{p}}^2}{2m_e}|
\Psi_{b\boldsymbol{k}'}\rangle,\\
H_0(b\boldsymbol{k},t\boldsymbol{k}')\!
=&\,2E_wS(b\boldsymbol{k},t\boldsymbol{k}')-
\langle \Psi_{b,\boldsymbol{k}}
|\frac{\hat{\boldsymbol{p}}^2}{2m_e}|
\Psi_{t,\boldsymbol{k}'}\rangle,
\end{align}
\end{subequations}
for different layers. The transformation $\tilde{C}=\sqrt{\mathbb{I}+\mathbb{S}}{C}$ reduces Eq. (\ref{eq:Matrix_Equation_general}) to $\hat{H}\tilde{C}=E\tilde{C}$,
\begin{align}\label{eq:H_general}
\hat{H}=
\frac{1}{\sqrt{\mathbb{I}+\mathbb{S}}}
\hat{H}_{0}
\frac{1}{\sqrt{\mathbb{I}+\mathbb{S}}}
\approx \hat{H}_{0}-
\frac{1}{2}
\left\{\hat{H}_0,\mathbb{S}
\right\},
\end{align}
where we expanded $(\mathbb{I}+\mathbb{S})^{-1/2}\approx\mathbb{I}-\frac{1}{2}\mathbb{S}$, and $\left\{\hat{X},\hat{Y}\right\}$ denotes the anticomutator of $\hat{X}$ and $\hat{Y}$.

The following two subsections are devoted to training the HkpTB approach with two systems, monolayer graphene and Bernal bilayer graphene. Using the known low-energy band structure, we quantify the input parameters in the HkpTB description and use them to describe the coupling across the twisted interface.
\subsection{Monolayer \textit{k}$\cdot$\textit{p} basis}
\label{sec:slg}

\begin{figure}
\begin{center}
\includegraphics[width=1\columnwidth]{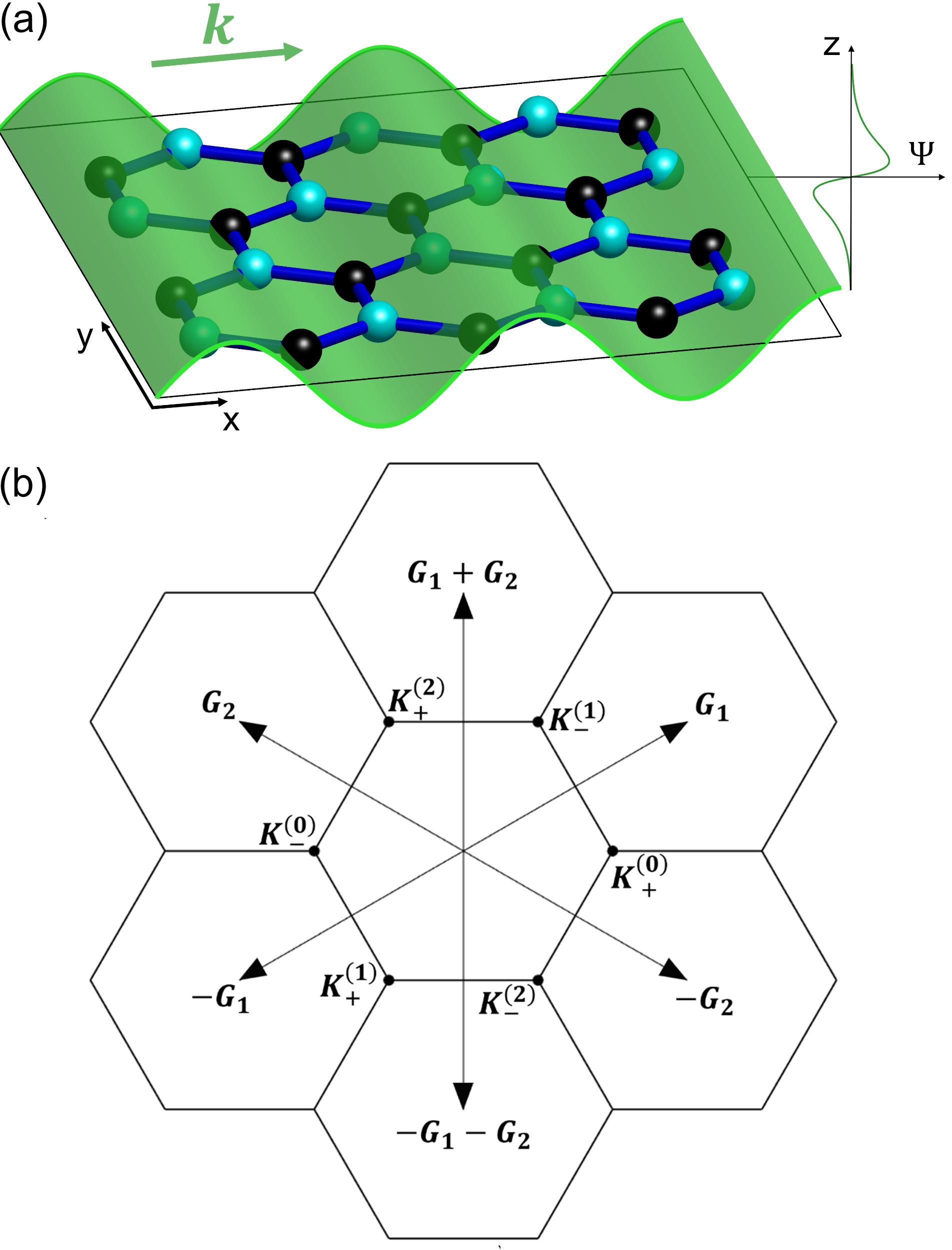}
\caption{ (a) Monolayer graphene lattice, with the A and B sublattices coloured in light blue and black, repectively. The eigenfunction $\Psi_{\boldsymbol{k}}$ is depicted by a plane-wave, with a z dependence amplitude inherited by $p_z$-orbitals of carbon atoms in graphene. (b) Graphene reciprocal space, highlighting the corners of the first Brillouin zone, $\boldsymbol{K}_\xi^{(j)}$, and six of the smallest reciprocal lattice vectors, $\boldsymbol{G}=\pm\boldsymbol{G}_1,\,\pm\boldsymbol{G}_2,\,\pm(\boldsymbol{G}_1+\boldsymbol{G}_2)$. \label{fig:MLG_BLG_and_BZ}}
\end{center}
\end{figure}

A monolayer graphene has carbon atoms sitting at the sites of a planar hexagonal lattice, which divides into two Bravais sublattices with a lattice constant of $a$ [see Fig. \ref{fig:MLG_BLG_and_BZ}(a)]. Around the corners of the hexagonal Brillouin zone, we use Bloch functions,
\begin{align}\label{eq:Bloch_TB}
\Psi_{\lambda,\boldsymbol{k}}^{(j)}(\boldsymbol{r},z)=&
\sum_{\boldsymbol{R}_\lambda}
\frac{
e^{i(\boldsymbol{K}_\xi^{(j)}+\boldsymbol{k})\cdot\boldsymbol{R}_\lambda} }{\sqrt{N}}
\Phi_{\lambda}(\boldsymbol{r}-\boldsymbol{R}_\lambda,z),
\\
\boldsymbol{R}_\lambda=&\,
n\boldsymbol{a}_1+m\boldsymbol{a}_2+\boldsymbol{\tau}_\lambda.\nonumber
\end{align}
In this expression, $\boldsymbol{a}_{1,2}=(\pm1/2,\sqrt{3}/2)a$ are the lattice vectors, $\lambda=A,B$ labels the sublattices, $\boldsymbol{\tau}_A = 0$, $\boldsymbol{\tau}_B = (0,a/\sqrt{3})$, $N$ is the number of unit cells and 
\begin{equation}
\boldsymbol{K}^{(j)}_\xi = \xi K [\cos(2\pi j/3), -\sin(2\pi j/3)],
\end{equation}
with $j = 0, 1, 2$, $\xi = \pm 1$ being the valley index and $K = 4\pi/3a$. In Eq. (\ref{eq:Bloch_TB}), we also introduce orbital functions $\Phi_\lambda$. Near the Fermi level, the energy bands primarily stem from the out-of-plane $p_z$ orbitals of the carbon atoms, affected by a trigonal directionality towards the closest neighbours on the honeycomb lattice. This is implemented in our theory by mixing the $s$-wave symmetric $p_z$ orbital with a small $f$-wave symmetric component,
\begin{equation}\label{eq:PzOrbitals}
\Phi_\lambda(\boldsymbol{r},z) = 
f(r,z) - \delta_\lambda r^3 
\sin(3\varphi)
\tilde{f}(r,z),
\end{equation}
where $\delta_A = \delta $, $\delta_B = - \delta$, and $\boldsymbol{r}=(x,y)=r(\cos\varphi,\sin\varphi)$ is the planar projection of the position vector.

We now expand the Bloch functions in Eq. (\ref{eq:Bloch_TB}) around $\boldsymbol{K}_\xi^{(0)}$. Retaining the slowly varying harmonics, it takes the form
\begin{align}\label{eq:Bloch_PW}
&\Psi_{\lambda,\xi,\boldsymbol{k}}^{(0)}(\boldsymbol{r},z)
\approx
\frac{e^{i\boldsymbol{k}\cdot\boldsymbol{r}}}{\sqrt{N}\Omega} \sum^2_{j=0} e^{i\boldsymbol{K}^{(j)}_\xi\cdot(\boldsymbol{r} - \tau_\lambda)} 
\mathcal{F}\left\{\Phi_\lambda\right\} (\boldsymbol{K}^{(j)}_\xi+\boldsymbol{k},z),
\end{align}
with $\Omega$ being the area of a unit cell and
\begin{align}\label{eq:Fourier}
&\mathcal{F}\left\{\Phi_\lambda\right\}
(\boldsymbol{K}^{(j)}_\xi+\boldsymbol{k},z)
\equiv
\int_{\mathbb{R}^2}
\mathrm{d}\boldsymbol{r}
e^{i(\boldsymbol{K}^{(j)}_\xi+\boldsymbol{k})\cdot\boldsymbol{r}}
\Phi_\lambda(\boldsymbol{r},z)\\
&=F(|\boldsymbol{K}^{(j)}_\xi+\boldsymbol{k}|^2,z)\nonumber\\
&\qquad-
\frac{\delta_\lambda}{2i}
\left[
\left(
\frac{\partial}{\partial k_x}+
i
\frac{\partial}{\partial k_y}
\right)^3-
\left(
\frac{\partial}{\partial k_x}-
i
\frac{\partial}{\partial k_y}
\right)^3
\right]
\tilde{F}(|\boldsymbol{K}^{(j)}_\xi+\boldsymbol{k}|^2,z).
\nonumber
\end{align}
Notice that the functions $F$ and $\tilde{F}$, the Fourier transform of $f$ and $\tilde{f}$ with respect to $\boldsymbol{r}$, respectively, are expressed as a function of the modulus square of the wavevector. This allows us to ease the functional form for the expansion around $\boldsymbol{K}_\xi^{(j)}$ of Eq. (\ref{eq:Fourier}), 
\begin{align}
\mathcal{F}\left\{\phi_s\right\}&
(\boldsymbol{K}_\xi^{(j)}+\boldsymbol{k},z)\approx
F(K^2,z)\\
&+2\boldsymbol{k}\cdot\boldsymbol{K}_\xi^{(j)}F'(K^2,z)
+i\delta_\lambda 24\xi K
\left(
\boldsymbol{k}\times\boldsymbol{K}_\xi^{(j)}
\right)_z
\tilde{F}^{'''}(K^2,z),\nonumber
\end{align}
where we used expansion in $ak\ll1$, and the prime symbol on $F$ and $\tilde{F}$ represents the derivative with respect to $K^2$. 

The orbital wave functions belonging to the same sub-lattice are orthonormalized, i.e., $\langle \Psi_{\lambda,\xi,\boldsymbol{k}}^{(0)}|\Psi_{\lambda,\xi',\boldsymbol{k}'}^{(0)} \rangle \approx \delta_{\boldsymbol{k},\boldsymbol{k}'} \delta_{\xi,\xi'}$. However, the overlap between orbital wave functions of different sublattices does not vanish. To the first order in $\boldsymbol{k}$, it is given by
\begin{align}
\langle \Psi_{A,\boldsymbol{k}}^{(0)}|\Psi_{B,\boldsymbol{k}}^{(0)}\rangle =&\,
\langle \Psi_{B,\boldsymbol{k}}^{(0)}|\Psi_{A,\boldsymbol{k}}^{(0)}\rangle^* \approx \,
s_0 \pi^*_\xi,\\
s_0 =& \int \frac{dz}{\Omega} F(K^2,z)F'(K^2,z),\nonumber 
\end{align}
with $\pi_\xi\equiv \xi k_x+ik_y$. To ease the notation, here and subsequently we omit the valley index $\xi$ in the subscript of Bloch functions, as it is a good quantum number. Terms proportional to $\delta$
have been left out since the orbital mixing is assumed small. In the Hilbert space spanned by $\Psi_{\lambda,\boldsymbol{k}}^{(0)}$, we introduce the overlap matrix analogue to that in Eq. (\ref{eq:Overlap_general}),
\begin{equation}\label{eq:Overlap_MLG}
S_{\lambda',\lambda}(\boldsymbol{k})\equiv\,
\langle \Psi_{\lambda,\boldsymbol{k}}^{(0)}|
\Psi_{\lambda',\boldsymbol{k}}^{(0)}\rangle -
\delta_{\lambda',\lambda},
\end{equation}
and the $\hat{H}_0$ matrix in Eq. (\ref{eq:schrodinger_monolayer}), with matrix elements
\begin{subequations}\label{eq:Ham_MLG}
\begin{align}
{H_0}_{\lambda',\lambda}(\boldsymbol{k}) \approx & \, E_w, \\
{H_0}_{B,A}(\boldsymbol{k}) = &\, 
{H^*_0}_{B,A}(\boldsymbol{k}) \approx \hbar v \pi^*_\xi,
\end{align}
\end{subequations}
where $v = E_ws_0/\hbar$ is the Dirac velocity. Here, we neglect intra-sublattice hopping. Strictly speaking, due to the non-vanishing overlap matrix $\mathbb{S}$, the band structure is not given by the eigenvalues of the matrix $\hat{H}_0$, but by those of $\hat{H}$. However, in this case, the anti-commutator in Eq. (\ref{eq:H_general}) contributes with a second order correction \cite{McCann_2013} in $\boldsymbol{k}$, which we neglect, leaving 
\begin{align}\label{eq:Ham_MLG}
\mathcal{H}_g(\boldsymbol{k}) =
\left(
\begin{matrix}
E_w&v\hbar\pi_\xi^*\\
v\hbar\pi_\xi&E_w
\end{matrix}
\right),
\end{align}
as Dirac Hamiltonian for monolayer graphene. 

\subsection{HkpTB model for aligned bilayer graphene with an arbitrary lateral interlayer off-set}
\label{sec:blg}

\begin{figure}
\begin{center}
\includegraphics[width=1\columnwidth]{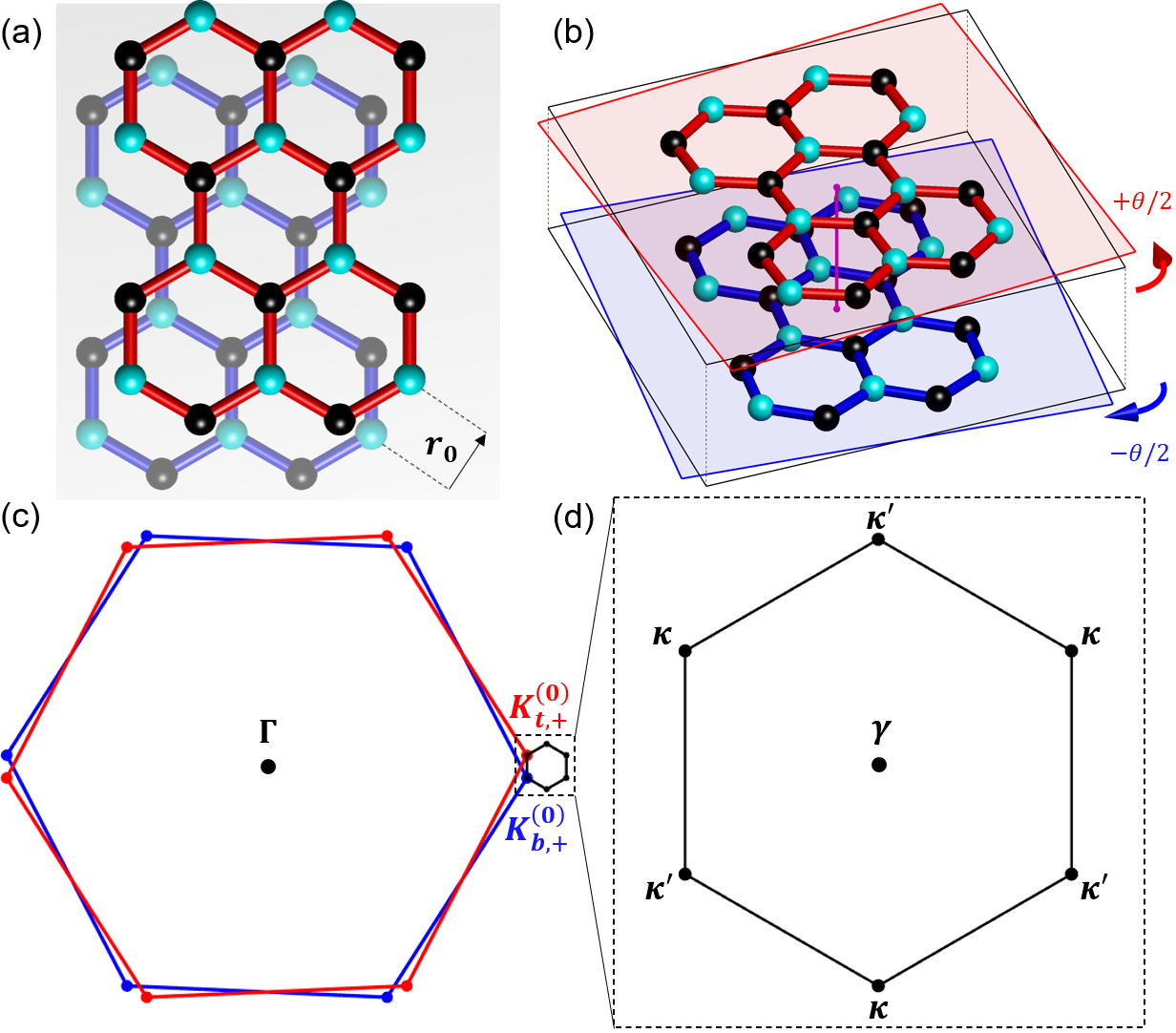}
\caption{ (a) Sketch of bilayer graphene, with a lateral offset $\boldsymbol{r}_0$ in the top layer, and (b) twisted bilayer graphene, with the top/bottom layer rotated by $\pm\theta/2$. The rotation axis is depicted by a purple vertical line. (c) First Brillouin zone of the top (bottom) in red (blue), highlighting the origin, or $\Gamma$-point, and the corner $\boldsymbol{K}_{t,+}^{(0)}$ ($\boldsymbol{K}_{b,+}^{(0)}$). (d) First moir\'{e} Brillouin zone (mBZ), highlighting the high symmetry points. \label{fig:BZ_and_mBZ}}
\end{center}
\end{figure}

Below, we consider aligned bilayer graphene: two layers rigidly stacked layers with the same crystallographic axes with a lateral offset $\boldsymbol{r}_0$, counted from AA stacking [see Fig. \ref{fig:BZ_and_mBZ} (a)]. In the present notation, the plane-wave representation of the Bloch functions for the bottom layer, denoted by $\Psi_{b,\lambda,\boldsymbol{k}}^{(0)}$, are identical to those of Eq. (\ref{eq:Bloch_PW}), while their counterparts in the top layer are
\begin{align}\label{eq:Bloch_PW_t}
\Psi_{t,\lambda,\boldsymbol{k}}^{(0)}(\boldsymbol{r},z) \approx &
\frac{e^{i(\boldsymbol{k}\cdot\boldsymbol{r}+\boldsymbol{K}_\xi^{(0)}\cdot\boldsymbol{r}_0)}}
{\sqrt{N}\Omega}\sum^2_{j=0}
e^{i\boldsymbol{K}^{(j)}_\xi\cdot(\boldsymbol{r} - \boldsymbol{\tau}_{\lambda}-\boldsymbol{r}_0)}\\
&\qquad\qquad\quad\times
\mathcal{F}\left\{\Phi_\lambda\right\}
(\boldsymbol{K}^{(j)}_\xi+\boldsymbol{k},z-c_0),\nonumber 
\end{align}
with $c_0$ being the interlayer distance. Bernal bilayer graphene corresponds to the special case $\boldsymbol{r}_0=\pm\boldsymbol{\tau}_B$, and AA stacking to $\boldsymbol{r}_0=0$. We analyse the form of $\hat{H}$ in Eq. (\ref{eq:H_general}) for Bernal stacking, with $\boldsymbol{r}_0=-\boldsymbol{\tau}_B$, and then use the band structure parameters of Bernal bilayers \cite{Slonczewski_1958, McClure_1957, McClure_1960} to train and parametrise the HkpTB model.

For the overlap matrix,
\begin{align} \label{eq:Overlap_BLG}
S_{l'\lambda',l\lambda}(\boldsymbol{k}) \equiv 
\langle \Psi_{l,\lambda,\boldsymbol{k}'}^{(0)}|\Psi_{l,\lambda,\boldsymbol{k}}^{(0)}\rangle - \delta_{\boldsymbol{k}',\boldsymbol{k}} \delta_{l',l}\delta_{\lambda',\lambda},
\end{align}
using the basis $\Psi^{(0)\,\dagger}_{\boldsymbol{k}} = \left(
\Psi^{(0)\,*}_{t,A,\boldsymbol{k}},
\Psi^{(0)\,*}_{t,B,\boldsymbol{k}},
\Psi^{(0)\,*}_{b,A,\boldsymbol{k}},
\Psi^{(0)\,*}_{b,B,\boldsymbol{k}}\right)$, the matrix elements of the interlayer sector read
\begin{subequations}
\begin{align}
S_{tA,bA}(\boldsymbol{k})=&\,
S_{tB,bB}(\boldsymbol{k})\\
=& \left[
6K\int \frac{\mathrm{d}z}{\Omega} F(K^2,z-c_0) F'(K^2,z)
\right]\pi^*_\xi,\nonumber\\
S_{tA,bB}(\boldsymbol{k})=& 
\left[
6K\int \frac{\mathrm{d}z}{\Omega} F(K^2,z-c_0) F'(K^2,z)\right.\\
&\left. - 72\delta K^2 \int \frac{\mathrm{d}z}{\Omega} F(K^2,z-c_0) \tilde{F}^{'''}(K^2,z)
\right]\pi_\xi\nonumber\\
S_{tB,bA}(\boldsymbol{k})=&\, 3\int \frac{\mathrm{d}z}{\Omega} 
F(K^2,z-c_0) F(K^2,z),
\end{align}
\end{subequations}
The equations above are then used to write the matrix elements of $\hat{H}$,
\begin{align}\label{eq:Ham_Bernal_BLG}
H_{t\lambda,b\lambda'}(\boldsymbol{k})=&
E_w S_{t\lambda,b\lambda'}(\boldsymbol{k}) + \frac{\hbar^2}{2m} \langle \Psi_{t,\lambda,\boldsymbol{k}}^{(0)}|\nabla^2|\Psi_{b,\lambda',\boldsymbol{k}}^{(0)}\rangle,\nonumber\\
\mathcal{H}(\boldsymbol{k}) \approx &
\begin{pmatrix}
\mathcal{H}_g(\boldsymbol{k})+\mathcal{E}_t&
T_{\boldsymbol{k}}\\
T_{\boldsymbol{k}}^\dagger&
\mathcal{H}_g(\boldsymbol{k})+\mathcal{E}_b
\end{pmatrix},\\ 
T_{\boldsymbol{k}}=&
\left(
\begin{matrix}
-\hbar v_4 \pi^*_\xi & -\hbar v_3 \pi_\xi \\
\gamma_1 & -\hbar v_4 \pi_\xi^*
\end{matrix}
\right), \nonumber\\
\mathcal{E}_\alpha=&
\left(
\begin{matrix}
\mathcal{E}_{\alpha A}&0\\
0&
\mathcal{E}_{\alpha B}
\end{matrix}
\right),\nonumber
\end{align}
where $\alpha=t,b$. The three parameters in $T_{\boldsymbol{k}}$ are related to the commonly used Bernal bilayer band structure parameters
\begin{subequations}\label{eq:BLG_paremeters}
\begin{align}
\gamma_1 =&
3\int\frac{dz}{\Omega} F(K^2,z-c_0) \left(E_w + \hat{O}(K,\partial_z)\right) F(K^2,z), \\
v_4 =&-\frac{6K}{\hbar} \int \frac{dz}{\Omega} F(K^2,z-c_0) \\
&\qquad\qquad\times\left[E_w + \hat{O}(K,\partial_z)\right] F'(K^2,z), \nonumber\\
v_3 =& v_4+\frac{72K^2 \delta}{\hbar} \int \frac{dz}{\Omega} F(K^2,z-c_0) \\
&\qquad\qquad\qquad\,\times\left[E_w + \hat{O}(K,\partial_z)\right] \tilde{F}^{'''}(K^2,z),\nonumber\\
&\hat{O}(Q,\partial_z) \equiv (\hbar^2/2m) \left(\partial^2_z - Q^2\right).\nonumber
\end{align}
\end{subequations}
The other commonly used parameter in Bernal bilayer graphene is the energy difference between dimer and non-dimer sites,
\begin{align}\label{eq:DeltaPrime}
\Delta'=\mathcal{E}_{tB}-\mathcal{E}_{tA}=\mathcal{E}_{bA}-\mathcal{E}_{bB}=9\mathcal{V},
\end{align}
where
\begin{align}\label{eq:Onsite_Energies}
\mathcal{E}_{t/b,\lambda}\equiv&\,
\langle\Psi_{t/b\lambda,\boldsymbol{k}'}^{(0)}|
V_{b/t}(\boldsymbol{r},z)
|\Psi_{t/b,\lambda,\boldsymbol{k}}^{(0)}\rangle,\\
\mathcal{E}_{tA}=&\,\mathcal{E}_{bB}=
-6\mathcal{V},\nonumber\\
\mathcal{E}_{tB}=&\mathcal{E}_{bA}=
3\mathcal{V},\nonumber\\
\mathcal{V}\equiv&
\frac{1}{\Omega}
\int_{-\infty}^{\infty}
\mathrm{d}zV_t^{\boldsymbol{G}}(z)
|F(K^2,z)|^2\nonumber\\
=&
\frac{1}{\Omega}
\int_{-\infty}^{\infty}
\mathrm{d}zV_b^{\boldsymbol{G}}(z)
|F(K^2,z-c_0)|^2,\nonumber\\
&V_{t/b}(\boldsymbol{r},z)=
\sum_{\boldsymbol{G}}
\left(
e^{i\boldsymbol{G}\cdot \boldsymbol{r}}+
e^{i\boldsymbol{G}\cdot (\boldsymbol{r}\pm\boldsymbol{\tau}_B)}
\right)
V_{t/b}^{\boldsymbol{G}}(z)\nonumber.
\end{align}
Here, the sum over $\boldsymbol{G}$ covers the six smallest reciprocal lattice vectors, with modulus $|\boldsymbol{G}|=4\pi/\sqrt{3}a$, as shown in Fig. \ref{fig:MLG_BLG_and_BZ} (b), and we omit the Fourier component of the potentials at $\boldsymbol{G}=\boldsymbol{0}$, which produces a constant diagonal term, absorbed in the overall energy shift of the resulting spectrum.

Equation~(\ref{eq:Ham_Bernal_BLG}) is the complete form of the SWMcC Hamiltonian, where we adopted the convention in Ref. \cite{Jeil_2014}, which includes a minus sign in front of $v_3>0$. This sign determines the orientation of the trigonal distortion and the three Dirac replicas of the band structure of bilayer graphene at very low energies, and it has been shown to be negative in DFT analysis \cite{Jeil_2014} and in recent ARPES measurements \cite{Joucken_2020}. We point out that an extra term in Eq.~(\ref{eq:BLG_paremeters}c), $(v_3-v_4)$, is a consequence of asymmetry of A and B on-site orbitals in Eq. (\ref{eq:PzOrbitals}).

For an arbitrary shift of the top layer $\boldsymbol{r}_0$, interlayer matrix elements of Hamiltonian $\hat{H}$ are
\begin{align}\label{eq:H_r0}
H_{tA,bA}(\boldsymbol{k})=&
H_{tB,bB}(\boldsymbol{k})\approx
\frac{\gamma_1}{3}
\sum_{j}
e^{i(\boldsymbol{K}_\xi^{(j)}-\boldsymbol{K}_\xi^{(0)})\cdot\boldsymbol{r}_0}\\
-&\frac{2}{3}
v_4\hbar
\sum_{j}
e^{i(\boldsymbol{K}_\xi^{(j)}-\boldsymbol{K}_\xi^{(0)})\cdot\boldsymbol{r}_0}
\boldsymbol{k}\cdot
\frac{\boldsymbol{K}_\xi^{(j)}}{K},\nonumber\\
%%%%%%%%%%%%%%%%%%
%%%%%%%%%%%%%%%%%%
%%%%%%%%%%%%%%%%%%
H_{tA,bB}(\boldsymbol{k}) 
\approx&
\frac{\gamma_1}{3}
\sum_{j}
e^{i(\boldsymbol{K}_\xi^{(j)}-\boldsymbol{K}_\xi^{(0)})\cdot\boldsymbol{r}_0}
e^{i\xi\frac{2\pi}{3}j}\nonumber\\
-&\frac{2}{3}
v_4\hbar
\sum_{j}
e^{i(\boldsymbol{K}_\xi^{(j)}-\boldsymbol{K}_\xi^{(0)})\cdot\boldsymbol{r}_0}
e^{i\xi\frac{2\pi}{3}j}
\boldsymbol{k}\cdot
\frac{\boldsymbol{K}_\xi^{(j)}}{K}
\nonumber\\
+&i\xi\frac{2}{3}(v_3-v_4)\hbar
\sum_{j}
e^{i(\boldsymbol{K}_\xi^{(j)}-\boldsymbol{K}_\xi^{(0)})\cdot\boldsymbol{r}_0}
e^{i\xi\frac{2\pi}{3}j}
\left(
\boldsymbol{k}\times
\frac{\boldsymbol{K}_\xi^{(j)}}{K}
\right)_z,\nonumber
\\
H_{tB,bA}(\boldsymbol{k}) 
\approx&
\frac{\gamma_1}{3}
\sum_{j}
e^{i(\boldsymbol{K}_\xi^{(j)}-\boldsymbol{K}_\xi^{(0)})\cdot\boldsymbol{r}_0}
e^{-i\xi\frac{2\pi}{3}j}\nonumber\\
-&\frac{2}{3}
v_4\hbar
\sum_{j}
e^{i(\boldsymbol{K}_\xi^{(j)}-\boldsymbol{K}_\xi^{(0)})\cdot\boldsymbol{r}_0}
e^{i\xi\frac{2\pi}{3}j}
\boldsymbol{k}\cdot
\frac{\boldsymbol{K}_\xi^{(j)}}{K}
\nonumber\\
-&i\xi\frac{2}{3}(v_3-v_4)\hbar
\sum_{j}
e^{i(\boldsymbol{K}_\xi^{(j)}-\boldsymbol{K}_\xi^{(0)})\cdot\boldsymbol{r}_0}
e^{-i\xi\frac{2\pi}{3}j}
\left(
\boldsymbol{k}\times
\frac{\boldsymbol{K}_\xi^{(j)}}{K}
\right)_z,\nonumber
\end{align}
while those in the intralayer sector are
\begin{align}\label{eq:Delta_r0}
\mathcal{E}_{tA}=&\,\mathcal{E}_{bB}=
\frac{\Delta'}{9}
\sum_{\boldsymbol{G}}
\left(
1+e^{-i\boldsymbol{G}\cdot\boldsymbol{\tau}_B}
\right)
e^{i\boldsymbol{G}\cdot\boldsymbol{r}_0},\\
\mathcal{E}_{tB}=&\,\mathcal{E}_{bA}=
\frac{\Delta'}{9}
\sum_{\boldsymbol{G}}
\left(
1+e^{+i\boldsymbol{G}\cdot\boldsymbol{\tau}_B}
\right)
e^{i\boldsymbol{G}\cdot\boldsymbol{r}_0}.\nonumber
\end{align}
For the case $\boldsymbol{r}_0=-\boldsymbol{\tau}_B$, we recover the matrix elements of Bernal bilayer graphene.

\begin{table*}
\begin{center}
 \begin{tabular}{||c | c | c | c | c| c | c | c ||}
 \hline
 Bilayer graphene & $v$ (m/s) & $\gamma_1$ (eV) & $v_3$ (m/s) & $v_4$ (m/s) &$\Delta'$ (eV)&$\gamma_2$ (eV)&$\gamma_5$ (eV)\\ [0.5ex] 
 \hline
 \hline
 Kuzmenko et al \cite{Kuzmenko_2009} & 1.02$\cdot10^6$ & 0.381 & 1.23$\cdot10^5$ & 4.54$\cdot10^4$ &0.022&-&-\\
 \hline
 Zhang et al \cite{Zhang_2008} & 3.0 & 0.40 & 0.3 & 0.15 &0.018&-&-\\
 \hline
 Jung et al \cite{Jeil_2014} & 8.45$\cdot10^5$ & 0.361 & 9.17$\cdot10^4$ & 4.47$\cdot10^4$ &0.015&-&-\\
 \hline
 \hline
 Bulk graphite & $v$ (m/s) & $\gamma_1$ (eV) & $v_3$ (m/s) & $v_4$ (m/s) &$\Delta'$ (eV)&$\gamma_2$ (eV)&$\gamma_5$ (eV)\\ [0.5ex] 
 \hline
 \hline
Dresselhaus et al \cite{Dresselhaus_2002} & 1.02$\cdot10^6$ & 0.39 & 1.02$\cdot10^5$ & 1.43$\cdot10^4$ &0.025&-0.020&0.038\\
 \hline
Yin et al \cite{Yin_2019} & 1.02$\cdot10^6$ & 0.39 & 1.02$\cdot10^5$ & 2.27$\cdot10^4$ &0.025&-0.017&0.038\\
 \hline
\end{tabular}
\end{center}
\caption{List of SWMcC parameters found in the literature. For numerical simulations on bilayer graphene, we adopt the parameters of Kuzmenko et al\cite{Kuzmenko_2009}. For numerical simlations of trilayer graphene, we adopt parameters by Dresselhaus et al.\cite{Dresselhaus_2002}. Relations of these parameters to the HkpTB parametrization used in the present work is given by Eqs. \ref{eq:BLG_paremeters}, \ref{eq:DeltaPrime}, \ref{eq:Onsite_Energies}, and \ref{eq:gamma_2and5}. Notice that, in our notation, the difference between dimer and non-dimer sites in bulk graphite is give by $2\Delta'$.  \label{table:SWMcC_parameters}}
\end{table*}

\section{Interlayer coupling across one twisted interface}
\label{sec:tblg}

In twisted bilayer graphene, the crystallographic axes of the two constituting layers form an angle $\theta$. For small angles, the Hamiltonian of one electron in such system can be constructed with the same methodology as in the previous section, but using a spatially modulated top-layer shift $\boldsymbol{r}_0=\theta\,\hat{\boldsymbol{z}}\times\boldsymbol{r}$, where $\hat{\boldsymbol{z}}$ is the unit vector in the vertical direction. As a result, the interlayer hybridisation acquires periodic coordinate dependence. For the minimal model, which corresponds to taking into account only the first term in each matrix element in Eq. (\ref{eq:H_r0}), the interlayer coupling becomes
\begin{align}
\left(
\begin{matrix}
H_{tA,bA}(\boldsymbol{r})&
H_{tA,bB}(\boldsymbol{r})\\
H_{tB,bA}(\boldsymbol{r})&
H_{tB,bB}(\boldsymbol{r})
\end{matrix}
\right)
\approx\!\frac{\gamma_1}{3}
\!\sum_{j=0}^2
\left(
\begin{matrix}
1&e^{i\xi\frac{2\pi}{3}j}\\
e^{-i\xi\frac{2\pi}{3}j}&1
\end{matrix}
\right)e^{i(\Delta\boldsymbol{K}_\xi^{(0)}-\Delta\boldsymbol{K}_\xi^{(j)})\cdot\boldsymbol{r}}.
\end{align}
Additionally, it is necessary to perform a unitary transformation to account for reciprocal space rotation between the top and bottom layers, which shifts their $K$-points, $\boldsymbol{K}_{t,\xi}^{(0)}$ in the top layer and $\boldsymbol{K}_{b,\xi}^{(0)}$ in the bottom layer, by $\pm\Delta\boldsymbol{K}_\xi^{(0)}/2$, respectively,
\begin{align}\label{eq:U}
&\mathcal{H}\to\mathcal{U}^\dagger\mathcal{H}\mathcal{U}\\
\mathcal{U}=&
\left(
\begin{matrix}
e^{i\Delta\boldsymbol{K}_\xi^{(0)}\cdot\boldsymbol{r}/2}
\mathbb{I}&0\\
0&
e^{-i\Delta\boldsymbol{K}_\xi^{(0)}\cdot\boldsymbol{r}/2}\mathbb{I}
\end{matrix}
\right),\nonumber
\end{align}
where $\mathbb{I}$ is the $2\times2$ unit matrix in the A-B sublattice space, and
\begin{align}\label{eq:DeltaK}
\boldsymbol{K}_\xi^{(j)}\cdot\boldsymbol{r}_0=
\theta\left(K_{\xi,y}^{(j)},-K_{\xi,x}^{(j)}\right)\cdot\boldsymbol{r}=
-\Delta\boldsymbol{K}_\xi^{(j)}\cdot\boldsymbol{r}.\nonumber
\end{align}
Unitary transformation in Eq. (\ref{eq:U}) gives a twisted interface coupling in the form \cite{Lopes_2007, Bistritzer_2011, Lopes_2012, Koshino_2015}
\begin{align}
\mathcal{H}_{\mathrm{MM}}(\boldsymbol{r})=&
\left(
\begin{matrix}
\mathcal{H}^t(\boldsymbol{r})&
\mathcal{T}(\boldsymbol{r})\\
\mathcal{T}^{\dagger}(\boldsymbol{r})&
\mathcal{H}^b(\boldsymbol{r})
\end{matrix}
\right),\\
\mathcal{H}^{t/b}(\boldsymbol{r})\equiv&
\left(
\begin{matrix}
0&v\hbar(-\xi i\partial_x-\partial_y\pm i\theta K/2)\\
v\hbar(-\xi i\partial_x+\partial_y\mp i\theta K/2)&0
\end{matrix}
\right),\nonumber\\
\mathcal{T}(\boldsymbol{r})
=&\frac{\gamma_1}{3}
\!\sum_{j=0}^2
\left(
\begin{matrix}
1&e^{i\xi\frac{2\pi}{3}j}\\
e^{-i\xi\frac{2\pi}{3}j}&1
\end{matrix}
\right)e^{-i\Delta\boldsymbol{K}_\xi^{(j)}\cdot\boldsymbol{r}}\nonumber
\end{align}

The HkpTB approach enables us to refine the description of the interlayer sector of the Hamiltonian, and the effect of modulating on-site energy, which is written below in the 2D plane-wave basis, suitable for the miniband analysis upon folding onto a small mini Brillouin zone of the moir\'{e} superlattice. The interlayer hybridisation of states is, then, described by
\begin{align}\label{eq:T_tBLG}
&\mathcal{T}_{\boldsymbol{k},\boldsymbol{k}'}=
\frac{\gamma_1}{3}
\sum_{j=0}^2
\left(
\begin{matrix}
1&e^{i\xi\frac{2\pi}{3}j}\\
e^{-i\xi\frac{2\pi}{3}j}&1
\end{matrix}
\right)
\delta_{\boldsymbol{k}',\boldsymbol{k}+
\Delta\boldsymbol{K}_\xi^{(0)}-
\Delta\boldsymbol{K}_\xi^{(j)}},\\
&\qquad+
\frac{\hbar v_4}{3K}
\sum_{j=0}^2(\boldsymbol{k}+\boldsymbol{k}')\cdot \boldsymbol{K}^{(j)}_\xi
\left(
\begin{matrix}
1&e^{i\xi\frac{2\pi}{3}j}\\
e^{-i\xi\frac{2\pi}{3}j}&1
\end{matrix}
\right)
\delta_{\boldsymbol{k}',\boldsymbol{k}+
\Delta\boldsymbol{K}_\xi^{(0)}-
\Delta\boldsymbol{K}_\xi^{(j)}}\nonumber\\
&\qquad+
i\xi \frac{\hbar (v_3-v_4)}{3K}
\sum_{j=0}^2
\left[(\boldsymbol{k}+\boldsymbol{k}')
\times\boldsymbol{K}^{(j)}_\xi\right]_z 
\left(
\begin{matrix}
0&e^{i\xi\frac{2\pi}{3}j}\\
-e^{-i\xi\frac{2\pi}{3}j}&0
\end{matrix}
\right)\nonumber\\
&\qquad\qquad\qquad\qquad\quad\times\delta_{\boldsymbol{k}',\boldsymbol{k}+
\Delta\boldsymbol{K}_\xi^{(0)}-
\Delta\boldsymbol{K}_\xi^{(j)}},\nonumber
\end{align}
supplemented by the potential created by one layer on the other,
\begin{align}\label{eq:E_tBLG}
&\mathcal{E}_{\boldsymbol{k}',\boldsymbol{k}}=
\frac{\Delta'}{9}
\sum_{\boldsymbol{G}}
\left(
\begin{matrix}
1+e^{i\boldsymbol{G}\cdot\boldsymbol{\tau}_B}&0\\
0&1+e^{-i\boldsymbol{G}\cdot\boldsymbol{\tau}_B}
\end{matrix}
\right)
\delta_{\boldsymbol{k}',\boldsymbol{k}+\Delta\boldsymbol{G}}.
\end{align}
Here, $\Delta\boldsymbol{G}\equiv\theta \hat{\boldsymbol{z}}\times\boldsymbol{G}$ is the wavenumber mismatch between the reciprocal lattice vector of the top and bottom layers, and we recall that the sum over $\boldsymbol{G}$ extends over the six smallest reciprocal lattice vectors, $\pm\boldsymbol{G}_1$, $\pm\boldsymbol{G}_2$ and $\pm(\boldsymbol{G}_1+\boldsymbol{G}_2)$.
\begin{figure*}
\begin{center}
\includegraphics[width=2\columnwidth]{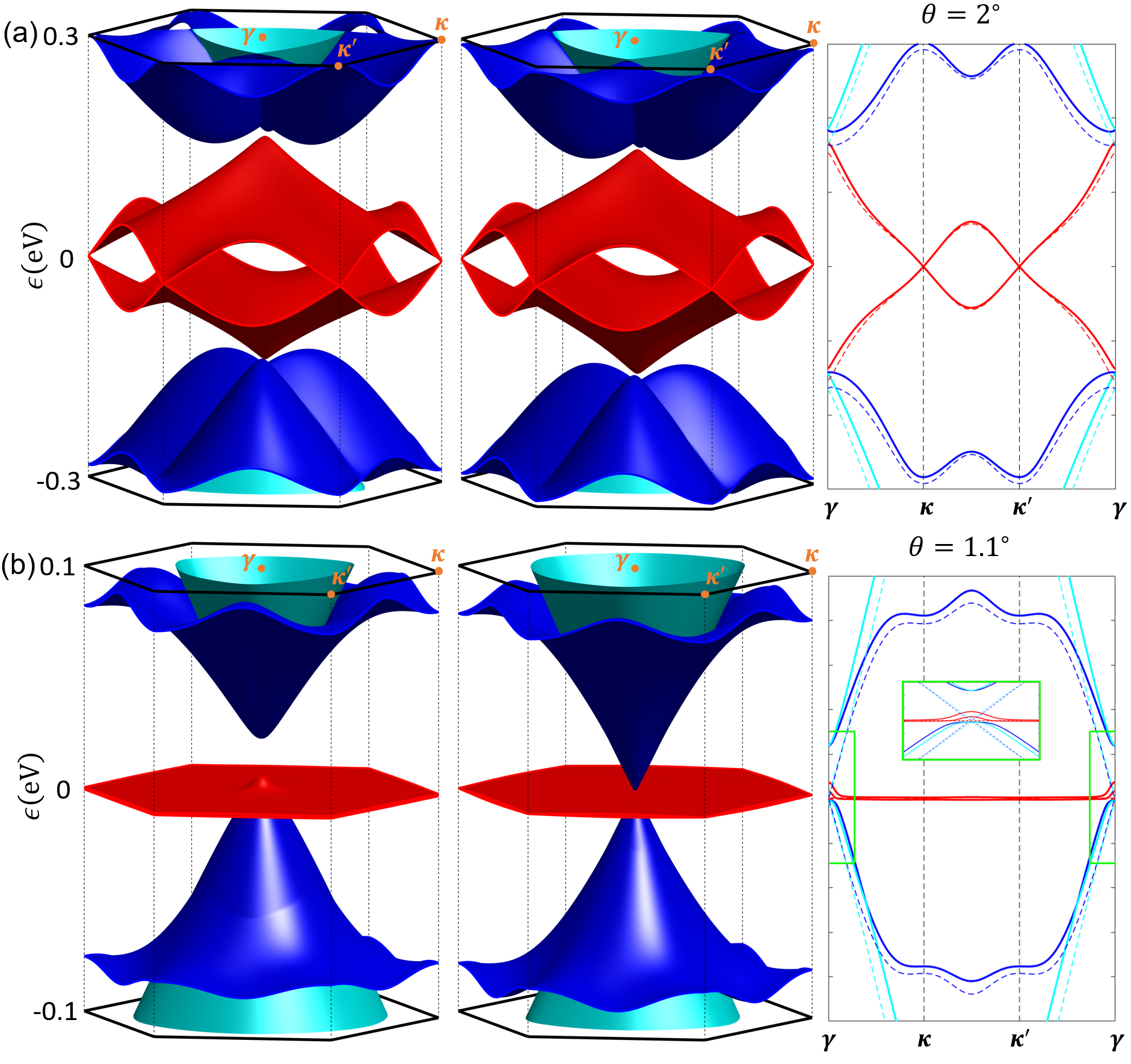}
\caption{Twisted bilayer graphene at (a) $\theta=2^\circ$ and (b) $\theta=1.1^\circ$, with $v=1.02\cdot10^6$ m/s, $\gamma_1=0.381$ eV, $v_3=1.23\cdot10^5$ m/s, $v_4=4.54\cdot10^4$ m/s and $\Delta'=0.022$ eV\cite{Kuzmenko_2009}. Band structure using the HkpTB model for the twisted interface (left) and band structure using the minimal model \cite{Lopes_2007,Bistritzer_2011}(middle). Band structure along the high symmetry cut $\boldsymbol{\gamma-\kappa-\kappa'-\gamma}$ using Eq. (\ref{eq:H_tBLG}) (solid lines) and minimal model (dashed lines) in the twisted interface. \label{fig:tBLG}}
\end{center}
\end{figure*}
\begin{figure*}
\begin{center}
\includegraphics[width=2\columnwidth]{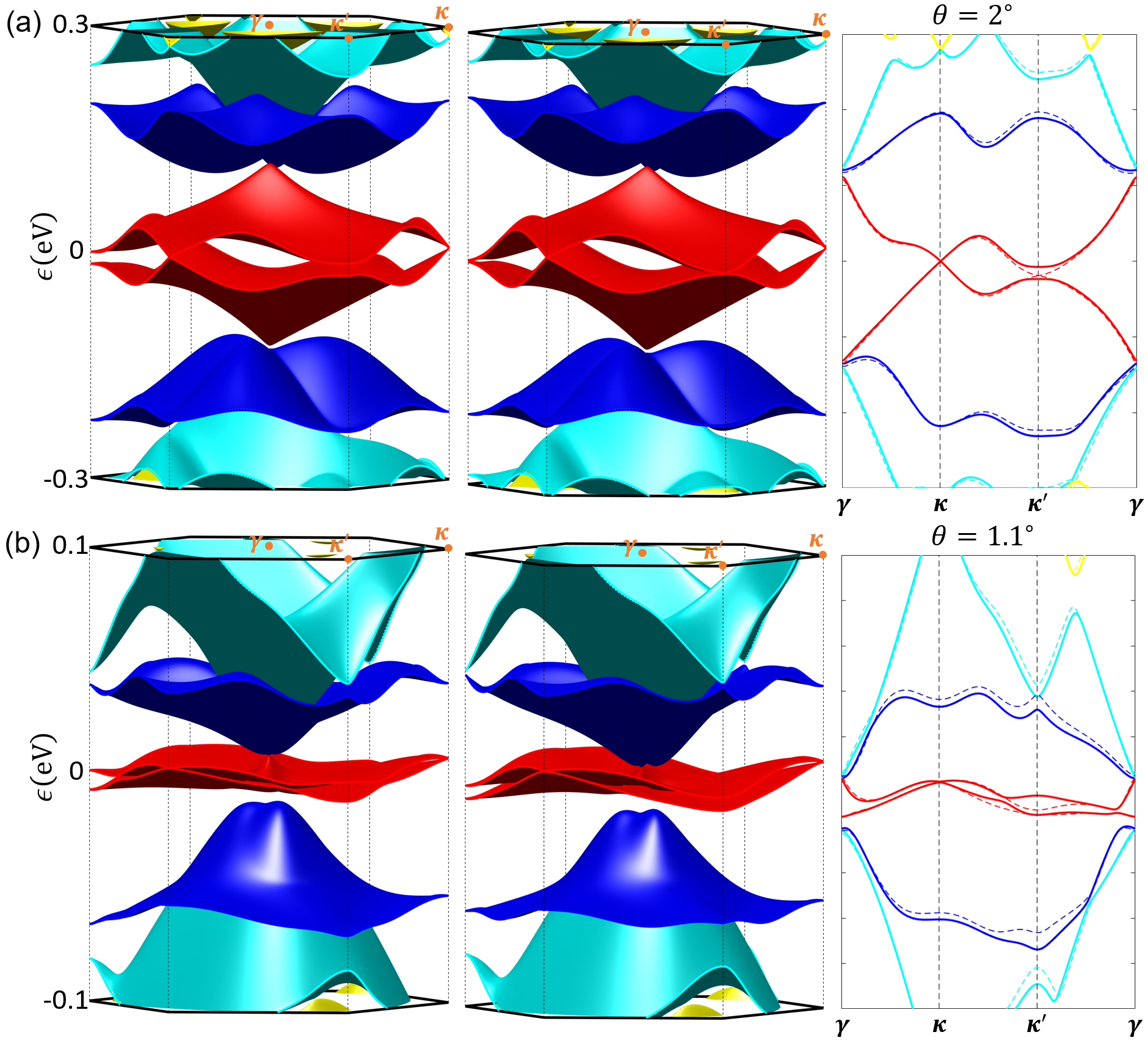}
\caption{Twisted trilayer graphene at (a) $\theta=2^\circ$ and (b) $\theta=1.1^\circ$, with $v=1.02\cdot10^6$ m/s, $\gamma_1=0.39$ eV, $v_3=1.02\cdot10^5$ m/s, $v_4=1.43\cdot10^4$ m/s, $\Delta'=0.025$ eV, $\gamma_2=-0.02$ eV, and $\gamma_5=0.038$ eV\cite{Dresselhaus_2002}. Band structure using the HkpTB model in the twisted interface (left), band structure using the minimal model (middle), and band structure along the high symmetry cut $\boldsymbol{\gamma-\kappa-\kappa'-\gamma}$ using the HkpTB (solid lines) and minimal (dashed lines) models in the twisted interface (right). \label{fig:tTLG}}
\end{center}
\end{figure*}

\section{Twisted bilayer graphene}
\label{sec:dispersion_tblg}
Now, we compute the miniband spectra of twisted bilayer using the refined twisted interface coupling in Eq. (\ref{eq:H_tBLG}) and compare it with the spectra computed using the minimal model. The miniband spectra are computed by zone folding and diagonalisation of plane-wave states coupled by the interlayer hybridization terms in Eq. (\ref{eq:T_tBLG}) and the additional moir\'{e} superlattice potential in Eq. (\ref{eq:E_tBLG}),

\begin{align}\label{eq:H_tBLG}
\mathcal{H}(\boldsymbol{k}',\boldsymbol{k})=&
\left(
\begin{matrix}
\mathcal{H}_{\boldsymbol{k}',\boldsymbol{k}}^t+
\mathcal{E}_{\boldsymbol{k}',\boldsymbol{k}}&
\mathcal{T}_{\boldsymbol{k}',\boldsymbol{k}}\\
\mathcal{T}_{\boldsymbol{k},\boldsymbol{k}'}^\dagger&
\mathcal{H}_{\boldsymbol{k}',\boldsymbol{k}}^b+
\mathcal{E}_{\boldsymbol{k}',\boldsymbol{k}}^{\dagger}
\end{matrix}
\right),\\
\mathcal{H}^{t/b}_{\boldsymbol{k}',\boldsymbol{k}}=&
\left(
\begin{matrix}
0&v\hbar\pi_{t/b,\xi}^{*}\\
v\hbar\pi_{t/b,\xi}&0
\end{matrix}
\right)
\delta_{\boldsymbol{k}',\boldsymbol{k}},\nonumber
\\
\pi_{t/b,\xi}=&\,\xi k_x+i(k_y\mp\theta K/2).
\nonumber
\end{align}

Examples of the resulting dispersions are shown in Fig. \ref{fig:tBLG} (a) and (b), for a larger angle and for a magic angle, respectively. For twist angles $\theta=2^\circ$, the band structure around the corners of the mini Brillouin zone, at $\boldsymbol{\kappa}$ and $\boldsymbol{\kappa}'$, inherits the conical dispersion from that of monolayer graphene, with a lower Dirac velocity \cite{Bistritzer_2011}. The band anticrossing produces saddle points in the first valence and conduction bands, which reflect themselves as van Hove singularities (vHS) in the density of states above and below the charge neutrality point \cite{Li_2010}. For $\theta = 2$ angle, the band structures obtained using the minimal model and full SWMcC HkpTB model are virtually indistinguishable.

The minimal model also predicts that the value for the renormalised Fermi velocity diminishes as the twist angle decreases\cite{Lopes_2007,Bistritzer_2011}, vanishing for an angle of $\theta=1.1^\circ$ [see Fig.\ref{fig:tBLG}(b)]. At this angle, the first valence and conduction bands span just a few meV, which results in a strong enhancement of the density of states. The additional terms introduced by the full set of SWMcC parameters in Eq. (\ref{eq:H_tBLG}) do not, does not change qualitatively this picture, yet they push other dispersive bands upwards in energy. This results in the formation of a gap of $\approx15$ meV on the conduction side, a clear spectral isolation of ``zero-energy" bands painted in red, from those above it in blue, and a small peak in the flat band dispersion around the moir\'{e} Brillouin zone $\gamma$-point, which agrees with the trend found in recent DFT calculations\cite{Fang_2016}. 

\section{Twisted trilayer (1+2) graphene}
\label{sec:dispersion_ttlg}
Here, we combine the generalised interlayer coupling in Eqs. (\ref{eq:T_tBLG}) and (\ref{eq:E_tBLG}) with the full SWMcC description of bilayer graphene in Eq. (\ref{eq:Ham_Bernal_BLG}) to decribe twisted trilayer (1+2) graphene: a monolayer stacked at a rotational fault upon a Bernal bilayer. As compared to twisted bilayers, for trilayer graphene, the SWMcC Hamiltonian contains an additional $2\times2$ blocks that accounts for tunnelling between the topmost layer and the bottommost layer\cite{Slonczewski_1958, McClure_1957, McClure_1960,Taychatanapat_2011}. In case of Bernal graphite, such couplings are accounted for by $\gamma_5$ and $\gamma_2$ hopping parameters, which distinguish the next-nearest layer coupling for electrons on the lower graphene sites that appear under the carbon (dimer sites with $\gamma_5$) or the empty center of hexagon (non-dimer sites with $\gamma_2$) in the layer above. The above mentioned difference reflects on the influence of the middle layer on the electron tunnelling between the outer layers in, e.g., Bernal trilayer. For a twistronic trilayer, we also account for such difference by distinguishing the couplings between A and B sublattice Bloch states in the bottom layer of Bernal bilayer with the plane wave states in the top (twisted) layer, which results in the overall trilayer Hamltonian,
\begin{align}
&\mathcal{H}(\boldsymbol{k'},\boldsymbol{k})=
\\
&\left(
\begin{matrix}
\mathcal{H}_{\boldsymbol{k}',\boldsymbol{k}}^t+
\mathcal{E}_{\boldsymbol{k}',\boldsymbol{k}}&
\mathcal{T}_{\boldsymbol{k}',\boldsymbol{k}}&
\mathcal{G}_{\boldsymbol{k}',\boldsymbol{k}}\\
\mathcal{T}_{\boldsymbol{k},\boldsymbol{k}'}^\dagger&
\mathcal{H}_{\boldsymbol{k}',\boldsymbol{k}}^b+
\mathcal{E}_{\boldsymbol{k}',\boldsymbol{k}}^{\dagger}+
\mathcal{E}_t&
T_{b,\boldsymbol{k}',\boldsymbol{k}}\\
\mathcal{G}_{\boldsymbol{k},\boldsymbol{k}'}^\dagger&T_{b,\boldsymbol{k}',\boldsymbol{k}}^\dagger&
\mathcal{H}_{\boldsymbol{k}',\boldsymbol{k}}^b+
\mathcal{E}_b
\end{matrix}
\right),\nonumber\\
&T_{b,\boldsymbol{k}',\boldsymbol{k}}\equiv
\left(
\begin{matrix}
-\hbar v_4 \pi^*_{b,\xi} & -\hbar v_3 \pi_{b,\xi} \\
\gamma_1 & -\hbar v_4 \pi_{b,\xi}^*
\end{matrix}
\right),\nonumber\\
&\mathcal{G}_{\boldsymbol{k}',\boldsymbol{k}}\equiv
\frac{1}{6}
\sum_{j=0}^2
\left(
\begin{matrix}
\gamma_{5}
e^{i\xi\frac{2\pi}{3}j}&
\gamma_{2}
e^{-i\xi\frac{2\pi}{3}j}\\
\gamma_{5}&
\gamma_{2}e^{i\xi\frac{2\pi}{3}j}
\end{matrix}
\right)
\delta_{\boldsymbol{k}',\boldsymbol{k}+
\Delta\boldsymbol{K}_\xi^{(0)}-
\Delta\boldsymbol{K}_\xi^{(j)}},\nonumber\\
\end{align}
where
\begin{align}\label{eq:gamma_2and5}
\gamma_{2,5}=&\,6\!\!\int\frac{dz}{\Omega} F(K^2,z-c_0)
\left[E_w + \hat{O}(K,\partial_z)\right]
F(K^2,z+c_0).
\end{align}
While the numerical values of these two parameters for twistronic trilayers, as well as the variations of on-site energies for the dimer and non-dimer sites, may differ from those in bulk graphite of Bernal bilayers, for example, due to a small variation of mean interlayer spacing, we expect their relative size to be similar. Hence, we use the values from Bernal graphite literature to assess the influence of these additional couplings on the miniband spectra of twistronic trilayers. 

The computed spectra of moir\'{e} minibands are shown in the leftmost panel of Figs. \ref{fig:tTLG} (a) and (b), for $\theta=2^\circ$ and $\theta=1.1^\circ$, respectively, in comparison with the minibands computed using the minimal model. The comparison of the two spectra shows that the influence of the additional terms, accounting for the full set of SWMcC couplings, is weak, suggesting that minimal model for the twisted interface coupling can be safely combined with the most detailed description of the Bernal-stacking part of twistronic few-layer graphene for the analysis of flat bands in such systems \cite{Polshyn_2020,Shi_2020,Garcia-Ruiz_2020}.

\section{Summary}
\label{sec:sum}
In this report, we construct the interacting Hamiltonian between two twisted graphene layers, first decomposing Bloch functions into plane-waves confined in the perpendicular direction, and then evaluating explicitly the matrix elements of single-particle Hamiltonian. We find that the zeroth-order expansion of the resulting coupling yields the well-known continuum model used in the literature. In turn, we find a contribution linear in momentum, which unlike existing models, recovers the skew coupling for the limiting case of $\theta\to0^\circ$ and bears new features to the band structure, such as the electron-hole asymmetry and the formation of gaps. For structures that combine both twisted and aligned interfaces, the first order contribution represents a small correction. This suggests that the simultaneous use of the minimal model for twisted interfaces and the full SWMcC description in the Bernal few-layer part of multilayer twistronic structure provides a reliable description of the band structure.
\section{Ackonowledgements}
This work was supported by European Graphene Flagship Core 3 Project, Lloyd Register Foundation Nanotechnology Grant, EC-FET Project 2D-SIPC, EPSRC grants EP/V007033/1, EP/S030719/1 and EP/N010345/1.

\end{document}